\documentclass[a4paper,11pt]{article}

\pdfoutput=1 % if your are submitting a pdflatex (i.e. if you have
\usepackage{jheppub} % for details on the use of the package, please
\usepackage[T1]{fontenc} % if needed
\usepackage{lmodern}

\usepackage{tikz}
\usetikzlibrary{positioning,decorations.pathmorphing}

\let\oldFootnote\footnote
\newcommand\nextToken\relax

\renewcommand\footnote[1]{%
    \oldFootnote{#1}\futurelet\nextToken\isFootnote}

\newcommand\isFootnote{%
    \ifx\footnote\nextToken\textsuperscript{,}\fi}

\usepackage{enumerate}

\usetikzlibrary{decorations.markings}%-------------------------------------%

\def\id{{1 \kern-.28em {\rm l}}}

\def\K3{{\bf K3}}
\def\journal#1&#2(#3){\unskip, \sl #1\ \bf #2 \rm(19#3) }
\def\andjournal#1&#2(#3){\sl #1~\bf #2 \rm (19#3) }

\def\bar{\overline}

\def\ie{{\it i.e.}}
\def\eg{{\it e.g.}}

\def\frac#1#2{{#1\over#2}}

\def\inbar{\,\vrule height1.5ex width.4pt depth0pt}
\def\IC{\relax\hbox{$\inbar\kern-.3em{\rm C}$}}
\def\IR{\relax{\rm I\kern-.18em R}}
\def\IP{\relax{\rm I\kern-.18em P}}

\catcode`\@=11
\def\slash#1{\mathord{\mathpalette\c@ncel{#1}}}
\def\EE{{\cal E}}
\def\FF{{\cal F}}

\def\NN{{\cal N}}

\def\SS{{\cal S}}

\def\underrel#1\over#2{\mathrel{\mathop{\kern\z@#1}\limits_{#2}}}

\catcode`\@=12

\def \sinh{{\rm sinh}}
\def \cosh{{\rm cosh}}

\def\exp{{\rm exp}}

\def\ie{{\it i.e.}}
\def\eg{{\it e.g.}}

\title{Comments on D3-Brane Holography}

\author{}
\author{Soumangsu Chakraborty$^a$,}
\author{Amit Giveon$^b$}
\author{and David Kutasov$^c$}

\affiliation[a]{Department of Theoretical Physics,\\Tata Institute of Fundamental Research, Mumbai 400005, India}
\affiliation[b]{Racah Institute of Physics, The Hebrew University \\ Jerusalem 91904, Israel}
\affiliation[c]{EFI and Department of Physics, University of Chicago\\ 5640 S. Ellis Ave, Chicago, IL 60637, USA}

\abstract{We revisit the idea that the quantum dynamics of open strings ending on $N$ D3-branes in the large $N$ limit can be described at large `t Hooft coupling by classical closed string theory in the background created by the D3-branes in asymptotically flat spacetime. We study the resulting thermodynamics and
compute the Hagedorn temperature and other properties of the D3-brane worldvolume theory in this regime. We also consider the theory in which the D3-branes are replaced by negative branes and show that its thermodynamics is well behaved. We comment on the idea that this theory can be thought of as an irrelevant deformation of $\NN=4$ SYM, and on its relation to $T\bar T$ deformed $CFT_2$.
}

\begin{document}
\maketitle
\flushbottom

\section{Introduction}\label{intro}

The correspondence between $\NN=4$ SYM with gauge group $SU(N)$ and gauge coupling $g_{YM}$ and string theory on $AdS_5\times S^5$, \cite{Maldacena:1997re,Gubser:1998bc,Witten:1998qj} (for a review, see \cite{Aharony:1999ti}),
can be motivated by studying a vacuum of type IIB string theory which contains $N$ coincident D3-branes,
and comparing two ways of thinking about it.

In one, we view it as a theory of closed strings propagating in all nine spatial dimensions, and open strings both of whose ends lie on the threebranes. At low energies, the open strings give rise to $\NN=4$ SYM with gauge coupling
 $g_{YM}^2=g$, the ten-dimensional string coupling, and the closed strings give rise to a $9+1$ dimensional gravitational theory.
 The two sectors decouple in the low-energy limit.

In the other, we view the D3-branes as a source of gravitational and other massless closed string fields,
and study the resulting background,
\begin{eqnarray}\label{d3}
ds^2&=&\frac{1}{\sqrt{f_3}}\left(-dt^2+\sum_{i=1}^3dx_i^2\right)+\sqrt{f_3}\left(dr^2+r^2d\Omega_{5}^2\right),\nonumber \\
e^{\Phi} &=& g,\\
%A_{01\cdots 3}&=& -\frac{(2\pi)^{3/2}l_s^4\lambda N}{f_3 r^4}.\nonumber
A_{01\cdots 3}&=& -\frac{r_3^4 N}{r^4f_3}.\nonumber
\end{eqnarray}
Here $(t,x_i)$ parametrize the D3-brane worldvolume, $r$ and $\Omega_5$ are spherical coordinates on the $\mathbb{R}^6$ transverse to the branes,
\begin{eqnarray}\label{harfun}
f_3=1+\frac{r_3^4}{r^{4}}=1+\frac{4 \pi\lambda l_s^4}{r^{4}},
\end{eqnarray}
and $\lambda$ is the 't Hooft coupling,
\begin{eqnarray}\label{thooft}
\lambda=g_{YM}^2N=gN.
\end{eqnarray}
From this point of view, the low-energy limit of the D3-brane worldvolume theory is described by the modes living in the near-horizon region
$r\ll r_3$, which is $AdS_5\times S^5$. The AdS radius, as well as the radius of the $S^5$, is given by $r_3$, \eqref{harfun}.
The decoupled closed strings live in the large $r$ region, which is asymptotically flat.

A natural question is whether it is possible to extend the above correspondence beyond the low-energy limit. A natural idea is to seek a correspondence between the worldvolume theory of the D3-branes and the full background \eqref{d3}, \eqref{harfun}.
This has the problem that for general $g$, $N$, the threebrane worldvolume theory is coupled to that of the closed strings,
so both of the above points of view describe the same theory -- the full string theory in a vacuum with $N$ D3-branes.
To get a non-trivial duality, we need to replace the low-energy limit, that gives rise to $\NN=4$ SYM and $AdS_5\times S^5$, by something else.

One idea \cite{Gubser:1998iu,Danielsson:2000ze,Kutasov:2001uf}  is to consider the theory in the limit
\begin{eqnarray}\label{declim}
g\to 0,\qquad N\to\infty,\qquad \lambda\;\;{\rm fixed}.
\end{eqnarray}
In this limit, the closed strings decouple, since the closed string coupling $g$ is sent to zero, while the open strings living on the D3-branes
remain interacting. From the worldsheet point of view, Riemann surfaces with additional handles are suppressed,
while those with additional holes contribute. As pointed out in \cite{Kutasov:2001uf}, this limit is reminiscent of the decoupling limit of NS5-branes
that leads to Little String Theory.

Thus, the idea is that the quantum theory of open strings ending on $N$ D3-branes, in the limit \eqref{declim}, is equivalent to classical closed string theory in the background \eqref{d3}, \eqref{harfun}. The latter, of course, contains the asymptotically flat region $r\to\infty$, and in particular, it contains all the closed string modes that live there. However, in the limit \eqref{declim}, these strings are free
and can be neglected for many purposes.\footnote{We will see later what this means in practice.}

Note also that the limit \eqref{declim} does not describe an extension of $\NN=4$ SYM with arbitrary $N$, $g_{YM}$ to finite energies, since it requires us to take the large $N$ limit. Rather, it trades the low-energy limit for the limit $N\to\infty$. The $1/N$ corrections to the resulting theory are an interesting question, which we will briefly comment on later.

At large $N$, the limit \eqref{declim} allows us to study physics at energies of order the string scale, $m_s=1/l_s=1/\sqrt{\alpha'}$,
for any $\lambda$, \eqref{thooft}. As is familiar from the AdS/CFT correspondence, the two descriptions of the dynamics are useful
in different regions in  coupling space.

For small $\lambda$, the useful description is in terms of open strings. Since each additional hole on the worldsheet comes with an additional power of $\lambda$, for small $\lambda$ we can do perturbative open string computations, while keeping only the lowest terms in the topological worldsheet expansion. The description \eqref{d3}, \eqref{harfun} is not useful in this regime, since the metric is strongly curved (on the string scale), except in the region $r\to\infty$, which as mentioned above describes the decoupled closed strings that we are not interested in.

For large $\lambda$, the open string description is not useful, since it involves computing the sum over worldsheets with an arbitrary number of holes (and perhaps non-perturbative effects). On the other hand, the closed string description \eqref{d3}, \eqref{harfun} is in this case weakly curved for all $r$, and thus we can use it to do calculations.

Another interesting part of the story has to do with the description of the theory as an irrelevant deformation of $\NN=4$ SYM . As mentioned above, as $r\to 0$ the background  \eqref{d3}, \eqref{harfun} approaches $AdS_5\times S^5$. Expanding around this limit gives an infinite series of corrections to the metric and other fields. The leading correction can be identified via the usual AdS/CFT map, \cite{Maldacena:1997re,Gubser:1998bc,Witten:1998qj}, as an addition to the Lagrangian of $\NN=4$ SYM of a dimension eight single-trace operator, that preserves $\NN=4$ SUSY but, of course, breaks conformal symmetry.

It has been proposed that the boundary theory dual to type IIB string theory in the background \eqref{d3}, \eqref{harfun} is such an irrelevant deformation of $\NN=4$ SYM \cite{Gubser:1998iu,Intriligator:1999ai,Danielsson:2000ze}; see also ~\cite{Gubser:1997yh,Gubser:1997se,Gubser:1998kv,Costa:1999sk,Costa:2000gk,Rastelli:2000xj,Evans:2001zn,Niarchos:2017cdz,Ferko:2019oyv,Caetano:2020ofu}. It is not clear to what extent this proposal is meaningful, because irrelevant deformations of conformal field theories are usually ill defined (since there is in general an infinite number of RG trajectories that approach a given irrelevant deformation of an IR CFT). However, in the last three years, a two-dimensional analog of the four-dimensional system discussed here was analyzed, and found to be better behaved than one might expect \cite{Giveon:2017nie,Giveon:2017myj,Asrat:2017tzd,Giribet:2017imm,Chakraborty:2018kpr,Chakraborty:2018aji}.

That system, often referred to as single-trace $T\bar T$ deformed CFT, is obtained by studying the near-horizon geometry of $k$ NS5-branes wrapped around $T^4\times S^1$ and $p$ fundamental strings wrapped around the $S^1$.  It gives rise to an $AdS_3\times S^3\times T^4$ vacuum of type IIB string theory, which is dual to a two-dimensional $(4,4)$ supersymmetric CFT with central charge $c=6kp$ \cite{Aharony:1999ti}.
Adding to the Lagrangian a certain single-trace irrelevant operator $D(x)$ \cite{Kutasov:1999xu} that preserves $(4,4)$ SUSY, gives rise in the bulk
to a background that interpolates between the geometry near the strings and one far from the strings (but still close to the fivebranes).

The string theory construction of \cite{Giveon:2017nie} provides some evidence for the claim that this particular irrelevant deformation of a CFT gives rise
to a well defined theory, that interpolates between a CFT in the IR and a theory with a Hagedorn spectrum (a particular two-dimensional vacuum of Little String Theory \cite{Aharony:1998ub}; for reviews, see  \cite{Aharony:1999ks,Kutasov:2001uf,Aharony:2004xn}) in the UV.

The purpose of this note is to reexamine the four-dimensional system \eqref{d3}, \eqref{harfun}, using tools that played a role
in the analysis of the two-dimensional one. We will focus on the thermodynamics of this system in the limit where the volume of
the threebranes $V_3$ is large (the thermodynamic limit). In this limit, the canonical free energy is expected to be dominated
by the contribution of a black hole in this background. We will analyze the resulting thermodynamics,
and discuss its relation to that of weakly coupled open strings ending on the threebranes.

Another goal of this note is to analyze the thermodynamics of the system \eqref{d3} with the harmonic function \eqref{harfun} replaced by
\begin{eqnarray}\label{harfunneg}
f_3=-1+\frac{r_3^4}{r^{4}}.
\end{eqnarray}
At first sight, this looks like an odd thing to do, however, in analogy to the two-dimensional situation \cite{Giveon:2017nie}, it can be thought of as adding the dimension eight
irrelevant deformation discussed above with an opposite sign coupling.\footnote{We will refer to this sign as negative.}

The background \eqref{d3}, \eqref{harfunneg} preserves the same $\NN=4$ SUSY as \eqref{d3}, \eqref{harfun}, but it is rather different from it.
In particular, it has a naked singularity at a finite value of the radial coordinate, $r=r_3$, and it is interesting to study its consequences.
It can be viewed as a four-dimensional analog of the negative coupling single-trace $T\bar T$ deformation of a $CFT_2$ discussed recently
from this point of view in \cite{toappear}. It is also related to the discussion of negative D-branes in \cite{Dijkgraaf:2016lym}.
We will see that, interestingly, it gives rise to a theory with apparently sensible thermodynamics.

\section{D3-brane thermodynamics}\label{sec2}

As explained in the previous section, for large $\lambda$ the D3-brane system is best described by the supergravity background \eqref{d3}, \eqref{harfun}.
In this section, we will study the canonical free energy of this background at a finite temperature $T$.  As usual, to do that we rotate time to Euclidean signature, and take it to be periodic with asymptotic period $\beta=1/T$. In the limit \eqref{declim}, the free energy receives contributions from a few backgrounds.

One is the original background with Euclidean time identified. This gives the free energy of closed strings in the background \eqref{d3}, \eqref{harfun}. Since this background is asymptotically flat $9+1$ dimensional spacetime, its contribution to the free energy goes like $F\sim V_9$, the volume of nine-dimensional space. We can write it as $F\sim V_3 V_6$, where $V_3$ is the volume of the space the D3-branes wrap, which we will take to be large but finite, and $V_6$ is the (infinite) transverse volume. The first factor is what we expect to get for the free energy of D3-branes in the thermodynamic limit, since the free energy is an extensive quantity. The second factor is infinite. This infinity is clearly due to the continuum of closed string modes with arbitrary momenta, propagating far from the threebranes. It has nothing to do with the threebranes, and in the limit \eqref{declim} the closed string modes that are responsible for it are decoupled. Therefore, we will neglect this contribution to the free energy.\footnote{This argument is actually a bit too quick. In principle, to compute the contribution of these closed strings, we should calculate the partition sum of closed strings in the background \eqref{d3} in the presence of a large $r$ regulator, subtract from it the regularized partition sum of closed strings in flat spacetime, and take the limit where the regulator goes to infinity. It is possible that in this limit a finite contribution remains, coming from the region in which the above two geometries differ, $r\sim r_3$. Such a contribution would be subleading in the limit \eqref{declim}, since it goes like $N^0$, whereas the other geometries discussed below contribute at order $N^2$.}

The second contribution to the free energy comes from non-extremal D3-branes. The corresponding Lorentzian background is given by \cite{Horowitz:1991cd}
\begin{eqnarray}\label{d3p}
ds^2&=&\frac{1}{\sqrt{f_3}}\left(-fdt^2+\sum_{i=1}^3dx_i^2\right)+\sqrt{f_3}\left(\frac{dr^2}{f}+r^2d\Omega_{5}^2\right),\nonumber \\
e^{\Phi} &=& g,\\
A_{01\cdots 3}&=& -\frac{1}{g}\left(\frac{r_0^4\sinh2\alpha}{2f_3 r^4}\right).\nonumber
\end{eqnarray}
The harmonic functions $f_3$, $f$ are given by
\begin{eqnarray}
f_3=1+\frac{r_3^{4}}{r^{4}}, \ \ \  f=1-\frac{r_0^{4}}{r^{4}}, \ \ \ r_3^4=r_0^4\sinh^2\alpha.\label{harfunp}
\end{eqnarray}
The geometry \eqref{d3p} has a horizon at $r=r_0$ and a singularity at $r=0$. The parameter $\alpha$ can be expressed in terms of the non-extremality parameter $r_0$,
\begin{eqnarray}\label{nd3p}
\lambda=\frac{1}{8\pi l_s^4 }r_0^4\sinh2\alpha.
\end{eqnarray}
In the limit $r_0\to 0$,  \eqref{d3p}, \eqref{harfunp} reduces to \eqref{d3}, \eqref{harfun}. Note that the radius $r_3$ in \eqref{harfunp} is not the same as in \eqref{harfun}.  In particular, it depends on $r_0$, $r_3=r_3(r_0)$. In the limit $r_0\to 0$, $r_3(r_0)\to r_3(0)=r_3$. We have omitted the dependence of $r_3$ on $r_0$ in \eqref{harfunp}, but it is important to remember that it's there.

\begin{figure}[h]
    \centering
    \includegraphics[width=.5\textwidth]{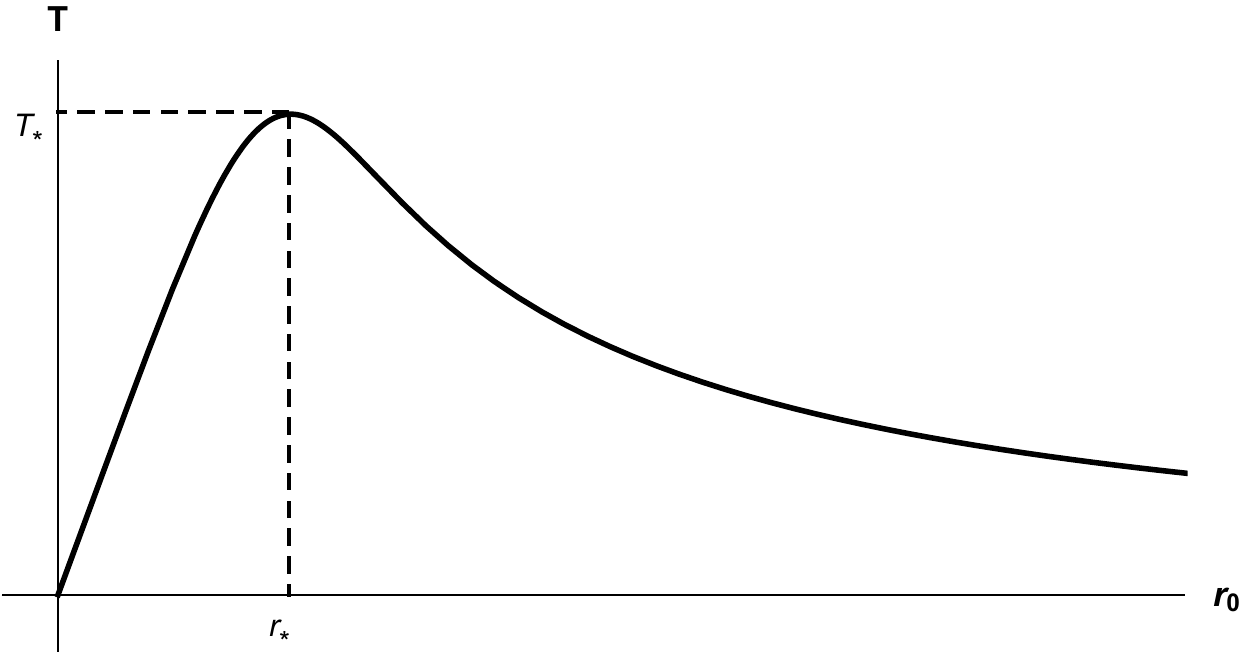}
    \caption{The Hawking temperature $T$ of a black hole as a function of its size $r_0$.}
    \label{img1}
\end{figure}

As usual, one can relate the non-extremality parameter $r_0$ to the temperature by compactifying Euclidean time on a circle of circumference $\beta$ and demanding that the resulting geometry is smooth. A short calculation leads to
\begin{eqnarray}\label{beta}
\beta=\frac{1}{T}=\pi r_0\cosh\alpha.
\end{eqnarray}
We plot the resulting relation in figure  \ref{img1}.

This figure has two prominent features:

\begin{enumerate}
\item For temperatures below a certain critical temperature $T_\ast$,
 \begin{eqnarray}
T_\ast=\frac{1}{2^{1/4}3^{3/8}\pi^{5/4}\lambda^{1/4} l_s} ,\label{tstar}
\end{eqnarray}
there are two black hole solutions, one with $r_0<r_*$, and the other with $r_0>r_*$,
\begin{eqnarray}
r_\ast^4=\frac{8}{\sqrt{3}}\pi \lambda l_s^4=\frac{2}{\sqrt{3}}[r_3(0)]^4.\label{rstar}
\end{eqnarray}
This is easy to understand: for low temperature, the horizon of the black hole with small $r_0$ is deep inside the $AdS_5\times S^5$ region in the geometry \eqref{d3}, \eqref{harfun}. Hence, it is a small deformation of the usual AdS-Schwarzschild black hole, that governs the thermodynamics of the low-energy theory ($\NN=4$ SYM at large `t Hooft coupling). On the other hand, the horizon of the black hole with large $r_0$ is deep inside the asymptotically flat region. Thus, it is a small deformation of a Schwarzschild black hole in flat $6+1$ dimensional spacetime. In order to determine which of these black holes dominates the thermodynamics at a particular temperature, one needs to compare their free energies, which we do below.
\item  As the temperature increases, the two black holes mentioned in (1) approach each other, and for $T=T_*$ they coincide (fig. \ref{img1}). For $T>T_*$ there are no black hole solutions. Thus, $T_*$ is a limiting temperature for the threebrane worldvolume theory. We will return to this issue later.
\end{enumerate}
Note that both $r_*$ and $\beta_*=1/T_*$ are comparable to $r_3(0)$, the AdS radius of the infrared $AdS_5\times S^5$ geometry, and the value of $r$ at which the $AdS_5$ throat connects to the asymptotically flat space far from the threebranes.

\begin{figure}[h]
    \centering
    \includegraphics[width=.5\textwidth]{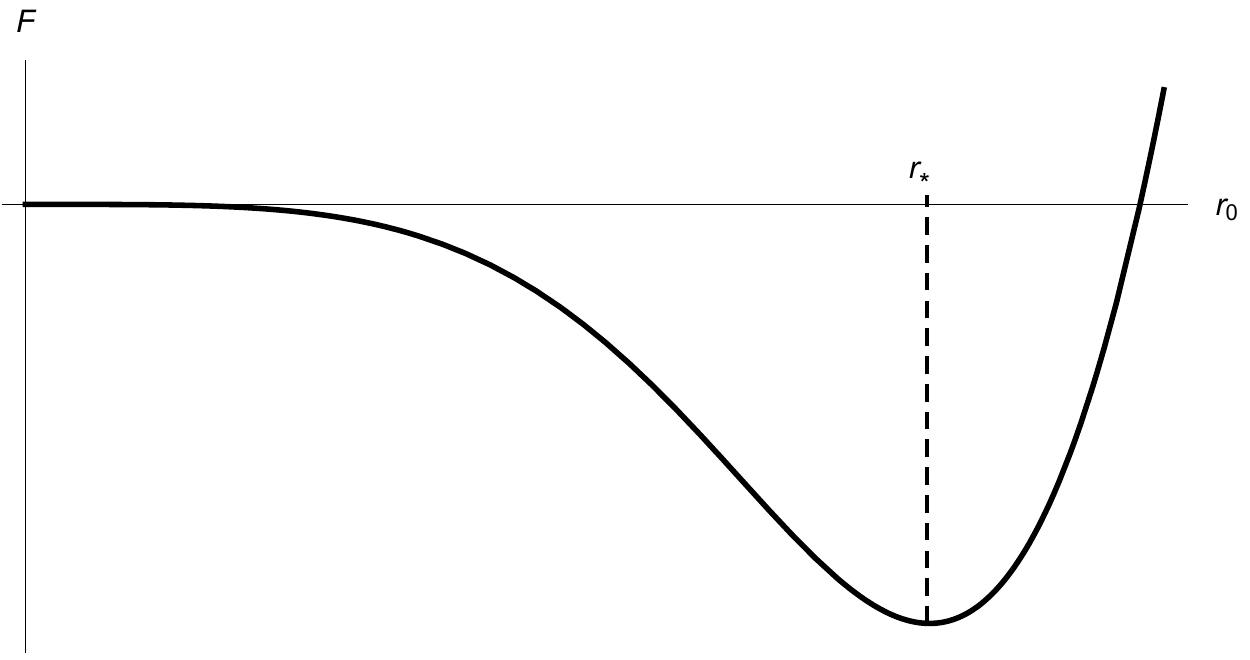}
    \caption{Free energy for non-extremal D3-branes.}
    \label{img2}
\end{figure}

To calculate the canonical free energy $F$, we use the thermodynamic relation
\begin{eqnarray}
F=E-TS.\label{deff}
\end{eqnarray}
In the thermodynamic limit, $F$, $E$, $S$ are extensive quantities, and it makes more sense to compute the corresponding densities. If we further take the large $N$ limit, these quantities are also proportional to $N^2$, the number of field theoretic degrees of freedom. Therefore, we will replace  \eqref{deff} by
\begin{eqnarray}
\FF=\EE-T\SS,\label{defff}
\end{eqnarray}
where
\begin{eqnarray}
\FF=\frac{F}{N^2V_3},\qquad \EE=\frac{E}{N^2V_3},\qquad \SS=\frac{S}{N^2V_3}.\label{deffff}
\end{eqnarray}
The entropy is given by the standard Bekenstein-Hawking formula,
\begin{eqnarray}\label{bhentp}
\SS=\frac{r_0^{5}\cosh\alpha }{32\pi^3\lambda^2l_s^8}.
\end{eqnarray}
To compute the energy, we start with the ADM mass,
\begin{eqnarray}\label{madmp}
M_{\rm ADM}=\frac{V_3N^2}{128\pi^4 \lambda^2l_s^8}r_0^{4}\left(3+2\cosh 2\alpha\right),
\end{eqnarray}
and subtract from it the extremal energy, $E_{\rm ext}$, the limit of \eqref{madmp} as $r_0\to 0$,
\begin{eqnarray}\label{engextp}
E_{\rm ext}=\frac{N^2 V_3}{(2\pi)^3 \lambda l_s^4}.
\end{eqnarray}
We get
\begin{eqnarray}\label{engaextdp}
\mathcal{E}=\frac{M_{ADM}-E_{ext}}{V_3N^2}=\frac{1}{128\pi^4 \lambda^2 l_s^8}\left(3r_0^4-16\pi \lambda l_s^4+2\sqrt{r_0^8+64\pi^{2}\lambda^2 l_s^8}\right).
\end{eqnarray}
Plugging \eqref{beta}, \eqref{bhentp}, \eqref{engaextdp} into \eqref{defff}, we get the free energy $\FF(r_0)$, plotted in figure \ref{img2}.

\begin{figure}[h]
    \centering
    \includegraphics[width=.5\textwidth]{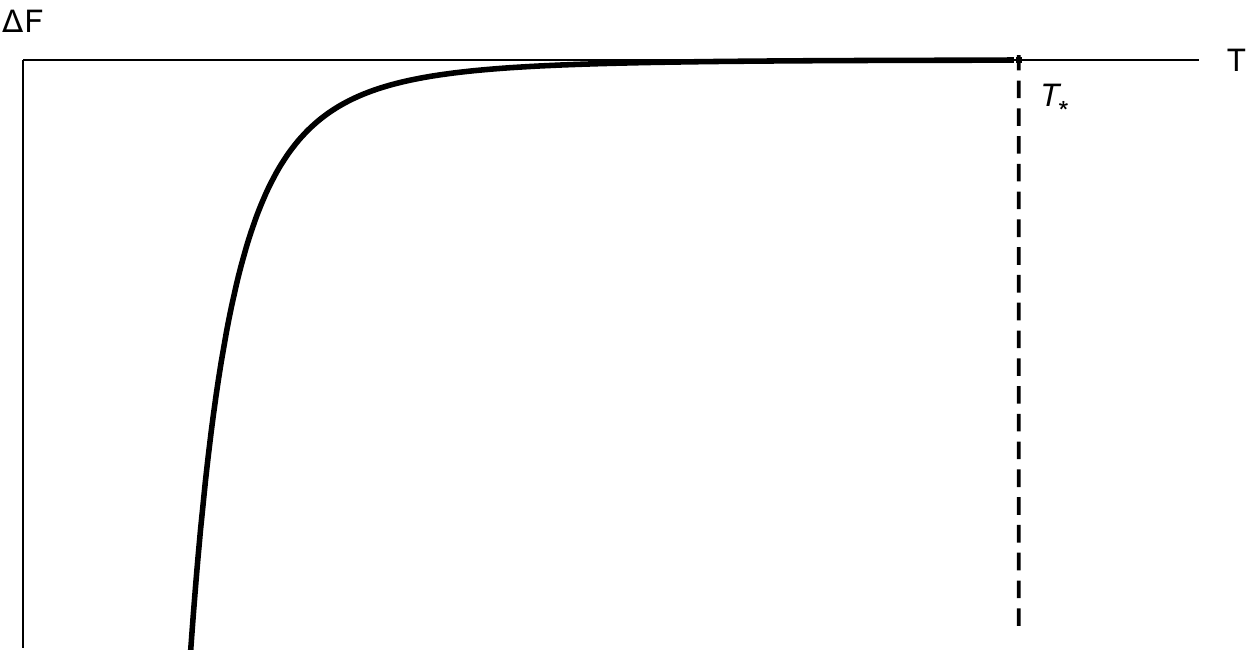}
    \caption{For all $T<T_*$, the small black hole has lower free energy than the large one with the same temperature $T$.}
    \label{img7}
\end{figure}

To compute the free energy as a function of temperature for $T<T_*$, we proceed as follows. Given the temperature, we can read off from figure \ref{img1} the two values of $r_0$ that correspond to it. Then we go to figure \ref{img2} and read off from it the difference of free energies between the small and large black holes, $\Delta\FF=\FF_{\rm small}-\FF_{\rm large}$. The result is plotted in figure \ref{img7}.

We see that the small black hole dominates the canonical ensemble for all $T<T_*$. Thus, in figure \ref{img2}, only the part of the curve to the left of the dashed line is physical, and it can be converted to a plot of $\FF(T)$ by solving for $r_0(T)$ using figure \ref{img1}.

It is interesting to study the behavior of the resulting free energy in various limits. At low temperature, $\FF$ has the following expansion
\begin{eqnarray}\label{freeexpt}
\FF&=&-\frac{\pi^2}{8}T^4\Big[1+2(\pi^{5}\lambda l_s^4 T^4)+8 (\pi^{5}\lambda l_s^4 T^4)^2+\cdots\Big].
\end{eqnarray}
We see that in the gravity approximation (large $\lambda$), the free energy is essentially a function of one parameter,
\begin{eqnarray}\label{genfree}
\FF&=&-\frac{\pi^2}{8}T^4\mathcal{G}(\lambda l_s^4T^4),
\end{eqnarray}
where $\mathcal{G}$ is a function whose power series expansion starts like \eqref{freeexpt}. The fact that in the gravity limit the non-trivial dependence is only on the parameter $\lambda l_s^4 T^4\simeq (r_3T)^4$ follows from a scaling argument \cite{Intriligator:1999ai}. Similarly to the two-dimensional case, one can think of $l_s^4$ as the coupling of the dimension eight operator in the deformed $\NN=4$ Lagrangian discussed in the introduction. Thus, in principle one should be able to get the expansion \eqref{freeexpt} from that field theory, although the field theory calculation (if it can be made sense of despite the fact that it represents a flow up the RG)  is likely easier in the weak coupling limit (small $\lambda$), whereas our results here are obtained at large $\lambda$.

\begin{figure}[h]
    \centering
    \includegraphics[width=.5\textwidth]{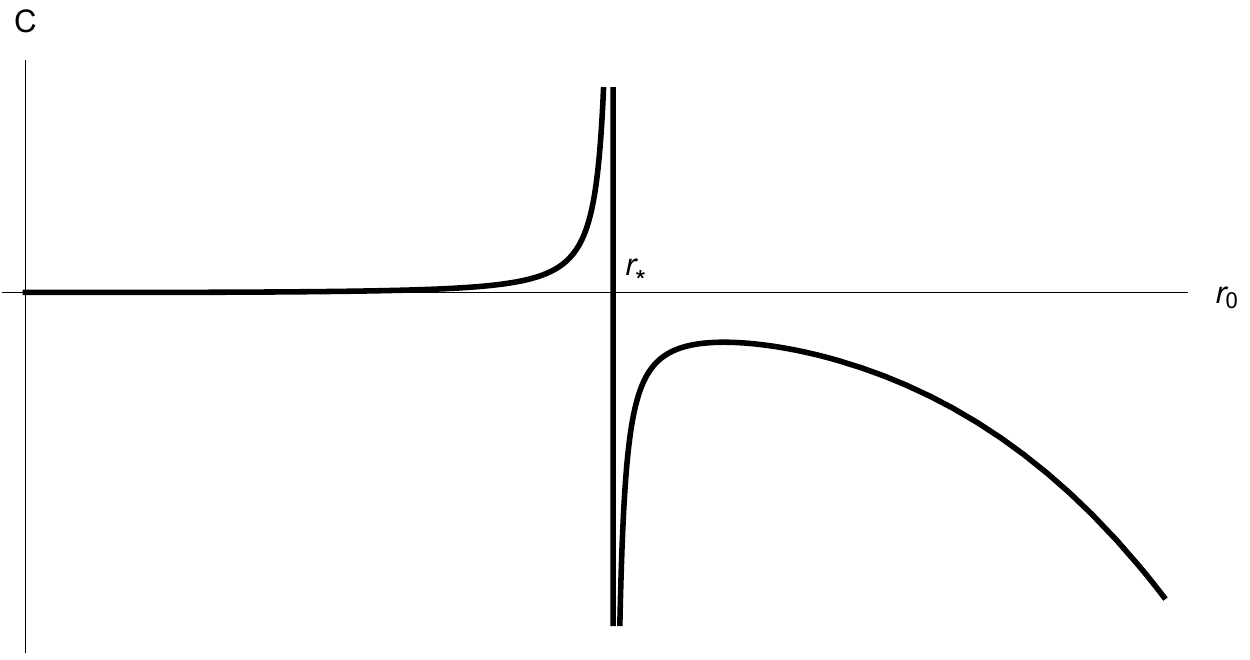}
    \caption{Specific heat $C$ vs $r_0$.}
    \label{img5}
\end{figure}

Another interesting limit is $T\to T_*$. By expanding the various quantities in $r_*-r_0$ (see appendix A) and eliminating $r_0$, we find that
\begin{eqnarray}\label{fengexp}
\FF=\FF_\ast+ \frac{\Delta T}{2^{7/4}3^{1/8}\pi^{7/4}\lambda^{3/4} l_s^3}-\frac{2^{15/8}\Delta T^{3/2}}{3^{23/16}\pi^{9/8} \lambda^{5/8} l_s^{5/2}}+O(\Delta T^2),
\end{eqnarray}
 where
 \begin{eqnarray}\label{fstar}
\FF_\ast=\frac{(\sqrt{3}-2)}{16\pi^{3}\lambda l_s^4},
\end{eqnarray}
 and
 \begin{eqnarray}\label{deltat}
 \Delta T=T_\ast-T.
 \end{eqnarray}
 We see that the free energy has a branch cut starting at $T=T_*$, but both the free energy and the energy density \eqref{engaextdp} are finite in the limit $T\to T_*$. The leading divergence is in the second derivative of $\FF$, \ie\ in the specific heat (see figure \ref{img5}).
Figure  \ref{img5} also implies that all black holes with $r_0<r_*$ have positive specific heat, while all those with $r_0>r_*$ have negative specific heat. This fits in well with the fact that for low temperature the former are to a good approximation  AdS-Schwarzschild black holes (which have positive specific heat), while the latter are approximately Schwarzschild black holes in flat spacetime (whose specific heat is negative).

 To recapitulate, for temperature $T<T_*$ \eqref{tstar} we find a sensible thermodynamics. The free energy $F(T)$ is described by the part of figures \ref{img1}, \ref{img2} with $r_0<r_*$. The specific heat is positive, see fig. \ref{img5}. The thermodynamics is dominated by black holes with $r_0<r_*$. One can think of them as  black holes that fit in the throat of the threebranes, the region in which the first term in the harmonic function $f_3$ \eqref{harfunp} is smaller than the second.

 As $T\to T_*$, the free energy develops a singularity \eqref{fengexp}. In particular, the specific heat diverges (figure \ref{img5}). The energy density \eqref{engaextdp} is bounded from above, $\EE<\EE_*$,
 \begin{eqnarray}\label{emax}
\EE_\ast=\frac{7\sqrt{3}-6}{48 \pi^3 \lambda l_s^4}.
\end{eqnarray}
 One can think of \eqref{emax} as the largest energy density that can fit into the throat of the threebranes. $T_*$ \eqref{tstar} is the maximal temperature for the threebranes. One can think of it as the Hagedorn temperature for open strings living on the threebranes at large $\lambda$.

 \section{Negative D3-brane thermodynamics}\label{sec3}

As mentioned in the introduction, one of the motivations for our analysis was the recent progress in $T\bar T$ deformed CFT \cite{Smirnov:2016lqw,Cavaglia:2016oda}, and in particular its single-trace version \cite{Giveon:2017nie,Giveon:2017myj,Asrat:2017tzd,Giribet:2017imm,Chakraborty:2018kpr,Chakraborty:2018aji}.
In that case, one can think of the theory as obtained by adding to the Lagrangian of a two-dimensional CFT holographically dual to string theory on $AdS_3\times S^3\times T^4$, say, an operator of dimension $(2,2)$, and following the RG to the UV. The physics of the resulting theory depends strongly on the sign of the coupling of this operator in the Lagrangian.

For positive sign, one gets in the UV a theory that has a Hagedorn density of states, reminiscent in some ways of the one discussed in section 2. For negative sign, one gets instead a theory with an upper bound on the energy, whose bulk description contains naked singularities, closed timelike curves, etc. Interestingly, for both signs the theory preserves $(4,4)$ SUSY, and in fact it is the theory with negative coupling that received more attention in the literature on $T\bar T$ deformed CFT.

In a recent paper \cite{toappear}, we studied the theory with negative coupling using tools similar to those employed here, and found that its thermodynamics appears to be sensible. It is an interesting open question whether this theory makes sense; \eg\ the work of \cite{Aharony:2018bad} suggests that its partition sum is not modular invariant.

The purpose of this section is to study the analogous problem for D3-branes. As discussed above, the addition of the dimension eight operator to the Lagrangian of $\NN=4$ SYM corresponds in the bulk to expanding the geometry \eqref{d3}, \eqref{harfun} around $r=0$. Thus, changing the sign of the deformation parameter is equivalent to changing the harmonic function $f_3$ from \eqref{harfun} to \eqref{harfunneg}, as in the two-dimensional case \cite{Giveon:2017nie,toappear}. This change seems problematic, since the resulting metric becomes complex for $r>r_3$. We will be able to avoid this issue, but it seems that to make sense of this theory, eventually one may need to address it.

Another thing to note is that the background  \eqref{d3}, \eqref{harfunneg} was studied from a different perspective in \cite{Dijkgraaf:2016lym}. These authors considered a background obtained by placing $N$ negative D3-branes, objects that preserve the same sixteen supercharges as the usual D3-branes, but have the opposite tension and Ramond charge, into a flat spacetime. Thus, for them the harmonic function $f_3$ was the negative of  \eqref{harfunneg}, and the signature of spacetime at large $r$ was the usual one $(9,1)$. Using a particular prescription for going to $r<r_3$, they argued that these negative D-branes are surrounded by a bubble where the signature of spacetime is different.

The main difference with our work is that from the point of view of \cite{Dijkgraaf:2016lym}, we are interested in the dynamics
``inside the bubble.'' Thus, we take the signature of the spacetime for $r<r_3$ to be the usual $(9,1)$,
and the time coordinate to be $t$ in \eqref{d3}.  From our perspective, the issue is what happens for $r>r_3$.

We now proceed to study the thermodynamics of the model \eqref{d3}, \eqref{harfunneg} following closely the discussion of section 2, and using \cite{toappear} as a guide. In this case, it is unclear what the contribution of the geometry with compactified Euclidean time is. In a sense, the singularity at $r=r_3$ provides a UV wall, so the contribution of perturbative closed string states in this geometry may not go like the infinite volume $V_6$. We leave this issue to future work.

We will assume that like for positive D3-branes, the contribution of perturbative strings in the geometry \eqref{d3}, \eqref{harfunneg} can be neglected, and the thermodynamics is governed by black holes in this background.  The analog of \eqref{d3p},\eqref{harfunp} for this case is
\begin{eqnarray}\label{d3n}
ds^2&=&\frac{1}{\sqrt{f_3}}\left(-fdt^2+\sum_{i=1}^3dx_i^2\right)+\sqrt{f_3}\left(\frac{dr^2}{f}+r^2d\Omega_{5}^2\right),\nonumber \\
e^{2\Phi} &=& g^2,\\
A_{01\cdots 3}&=& \frac{1}{g}\left(\frac{r_0^4\sinh2\alpha}{2f_3 r^4}\right),\nonumber
\end{eqnarray}
where
\begin{eqnarray}\label{harfunn}
f_3=-1+\frac{r_3^4}{r^{4}}, \ \ \  f=1-\frac{r_0^{4}}{r^{4}}, \ \ \ r_3^4=r_0^4\cosh^2\alpha.
\end{eqnarray}
$\alpha$ is related to $r_0$ via \eqref{nd3p}, as before.

The background \eqref{d3n}, \eqref{harfunn} is related to \eqref{d3p}, \eqref{harfunp} by $\alpha\to -\alpha+i\pi/2 $ and $(t,x_i)\to i(t,x_i)$. As in section 2, $r_3$, which in the present case is the location of a singularity, depends on $r_0$, $r_3=r_3(r_0)$. The last equation in \eqref{harfunn} implies that the singularity is always outside the black hole, $r_0<r_3(r_0)$, although $r_0$ can be much larger than the location of the singularity in the ground state, $r_3(0)$. All this is very similar to what we found in the two-dimensional case \cite{toappear}.

\begin{figure}[h]
    \centering
    \includegraphics[width=.5\textwidth]{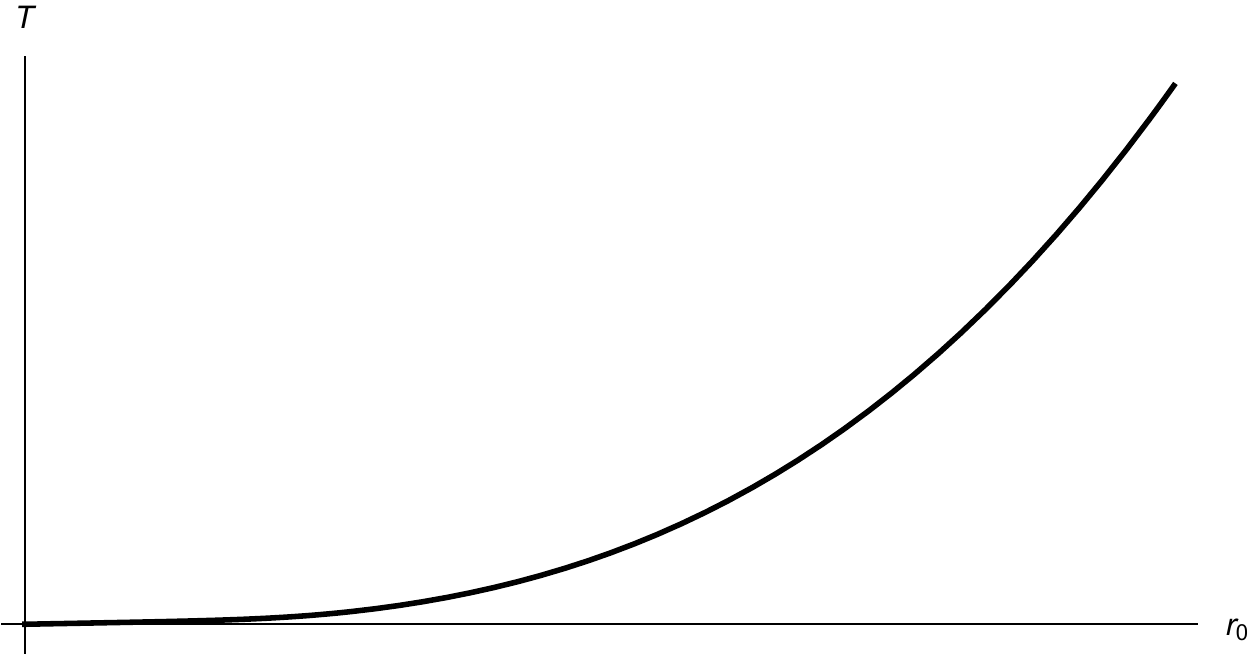}
    \caption{$T(r_0)$ for negative D3-branes.}
    \label{img3}
\end{figure}

Because of the fact that the singularity is always outside the horizon of the black hole, we can use the techniques of section 2 to study its thermodynamics.
The inverse temperature is given by
\begin{eqnarray}\label{betan}
\beta=\frac{1}{T}=\pi r_0\sinh\alpha.
\end{eqnarray}
 The temperature (plotted in figure \ref{img3}) increases monotonically from $0$ at $r_0=0$ and diverges as $r_0\to\infty$.

The Bekenstein-Hawking entropy  \eqref{deffff} is given by
\begin{eqnarray}\label{bhentdn}
\SS=\frac{1}{32\pi^3\lambda^2l_s^8}r_0^{5}\sinh\alpha.
\end{eqnarray}

\begin{figure}[h]
    \centering
    \includegraphics[width=.5\textwidth]{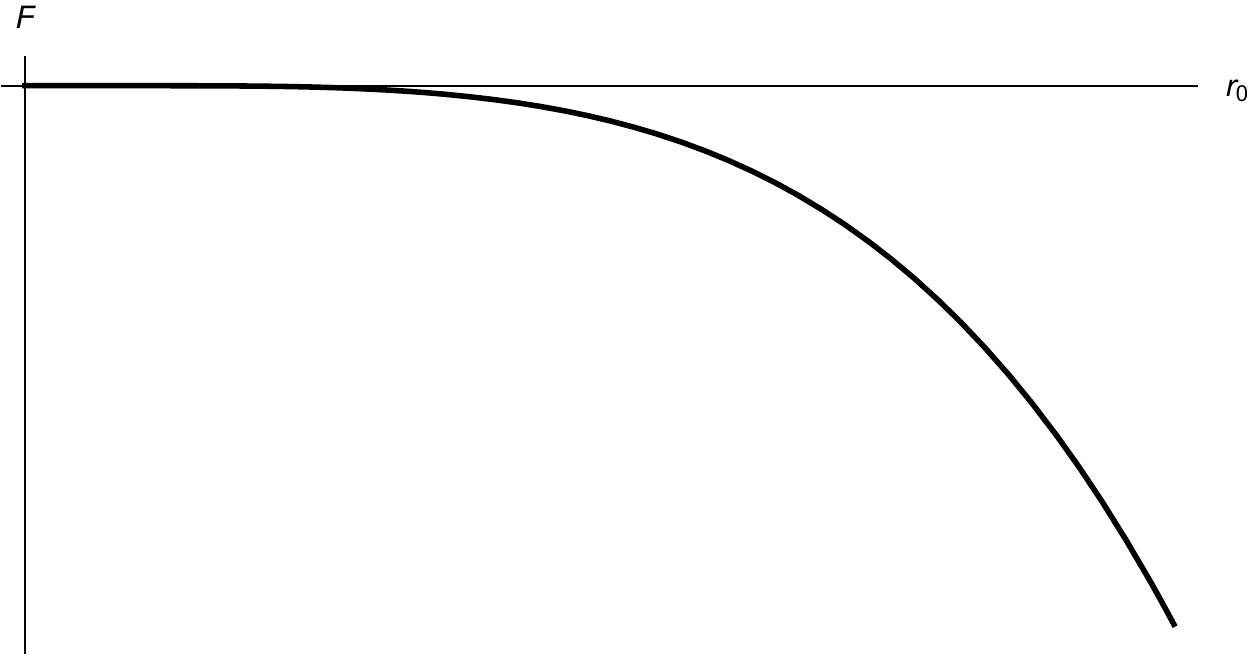}
    \caption{Free energy for negative D3-branes.}
    \label{img4}
\end{figure}

 The ADM mass  of the system is given by
\begin{eqnarray}\label{madmn}
M_{ADM}=\frac{V_3N^2}{128\pi^4 \lambda^2l_s^8}r_0^{4}\left(3-2\cosh 2\alpha\right).
\end{eqnarray}
As $r_0\to 0$, $M_{ADM}$ approaches
\begin{eqnarray}\label{engextpneg}
E_{\rm ext}=-\frac{N^2 V_3}{(2\pi)^3 \lambda l_s^4},
\end{eqnarray}
the negative of \eqref{engextp}. The energy above extremality is given by
\begin{eqnarray}\label{engaextdn}
\mathcal{E}=\frac{M_{ADM}-E_{ext}}{V_3N^2}=\frac{1}{128\pi^4\lambda^2 l_s^8}\left(3r_0^4+16\pi \lambda l_s^4-2\sqrt{r_0^8+64\pi^{2}\lambda^2 l_s^8}\right).
\end{eqnarray}
Plugging \eqref{betan}, \eqref{bhentdn}, \eqref{engaextdn} into \eqref{defff}, we get the free energy plotted in figure \ref{img4}.

\noindent
Interestingly, in this case, which one may have thought should be worse behaved than the one in section 2 (based on the geometry in the two cases), the thermodynamics is actually in some ways better behaved. Some of its properties are

\begin{figure}[h]
    \centering
    \includegraphics[width=.5\textwidth]{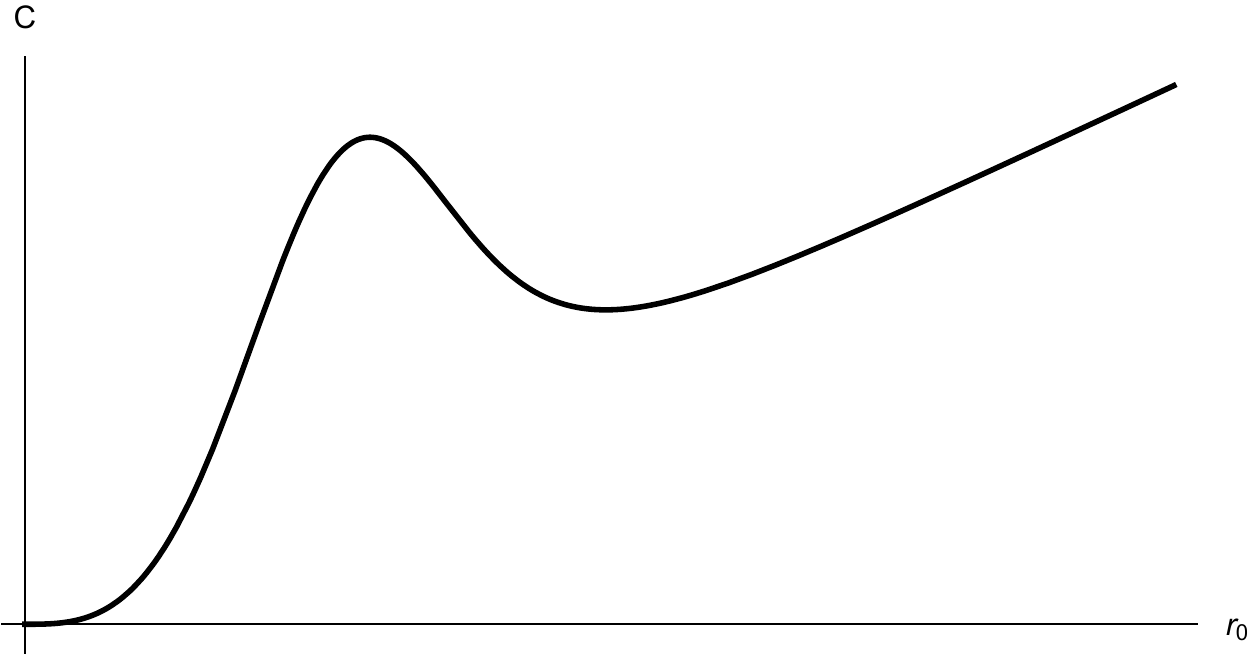}
    \caption{Specific heat for negative branes.}
    \label{img6}
\end{figure}

\begin{enumerate}
\item There is a unique black hole solution for all $0<T<\infty$ (figure \ref{img3}).
\item The specific heat is positive for all $T$ (figure \ref{img6}).
\item There is no upper bound on the energy and entropy density \eqref{deffff}. For large $\EE$, one has
 \begin{eqnarray}\label{entbhflatn}
 \SS=\frac{1}{2^{5/4}\pi \sqrt{\lambda} l_s^2}\EE^{1/4}+\cdots
 \end{eqnarray}
 \item The free energy density  $\FF$ has the low-temperature expansion
 \begin{eqnarray}\label{freeexptn}
\FF&=&-\frac{\pi^2}{8}T^4\Big(1-2(\pi^{5} \lambda l_s^4 T^4)+8 (\pi^{5} \lambda l_s^4 T^4)^2+\cdots\Big).
\end{eqnarray}
Note that as expected, it is related to  \eqref{freeexpt} by $l_s^4\to-l_s^4$ (see also appendix A). This is consistent with the fact that from the low-energy point of view one can think of $l_s^4$ as the coupling of the aforementioned dimension eight operator in the perturbed $\NN=4$ SYM Lagrangian.
 \end{enumerate}

\section{Discussion}\label{dissc}

The main goal of this note was to study the thermodynamics of open strings ending on a stack of $N$ D3-branes in the double scaling limit \eqref{declim}. We
argued that at large $\lambda$ the leading contribution to the canonical free energy is due to the black brane background \eqref{d3p}, \eqref{harfunp}. The resulting free energy exhibits Hagedorn behavior, in the sense that there is a maximal temperature, $T_*$ \eqref{tstar}. As $T\to T_*$, the free energy develops a branch cut, \eqref{fengexp}, and the specific heat diverges (see figure \ref{img5} and appendix A). The energy density above extremality is bounded from above by $\EE_*$ \eqref{emax}.

It is natural to compare these results to what one expects at weak coupling. At $\lambda=0$, the spectrum of (free) open strings ending on D3-branes exhibits a Hagedorn temperature, $T_H\sim m_s$. Our results suggest that at strong coupling this temperature decreases to $T_*\sim m_s\lambda^{-\frac{1}{4}}\sim 1/r_3$, the inverse AdS radius.

It is also interesting to study the free energy for small value of the dimensionless parameter $Tl_s$. At strong coupling, we found that the free energy has a Taylor expansion in $\lambda(Tl_s)^4$, \eqref{freeexpt}. In the free theory, the free energy is a sum of contributions of massless and massive open string modes. The massless modes are free $\NN=4$ SYM, and give a contribution similar to the leading term in \eqref{freeexpt}, apart from the famous factor of $3/4$ due to the difference in the free energy between small and large `t Hooft coupling \cite{Gubser:1996de}. The massive modes make contributions proportional to their Boltzmann factors
$\exp(-\beta H)$, which go like powers of $\exp(-m_s/T)$. These contributions are non-perturbative in the expansion parameter $Tl_s$.

Thus, the analogs of the perturbative terms in the expansion \eqref{freeexpt} all vanish at zero coupling. Interestingly, this is also what one would deduce by setting $\lambda=0$ in  \eqref{freeexpt}. Doing this is of course unjustified, since  \eqref{freeexpt} was obtained at large $\lambda$, but it is not unprecedented. The leading term in the expansion of the free energy does depend on $\lambda$, but it approaches (different) constants at $\lambda\to 0,\infty$. Perhaps something similar happens to the subleading terms in the expansion  \eqref{freeexpt} as well.

It is natural to ask how to compute the free energy of the open strings ending on the D3-branes at small but non-zero coupling. The masses of the excited string modes might change slightly due to the interactions, but their contributions would presumably still be non-perturbative in $Tl_s$. The perturbative contributions to the expansion should be due to higher order $(\alpha'$) corrections to the Lagrangian of the massless modes, the $\NN=4$ SYM Lagrangian. The leading such correction is the dimension eight operator mentioned in the introduction. Its coupling in the Lagrangian is proportional to $l_s^4$; thus, we expect its leading contribution to the free energy to go like $(l_sT)^4$, precisely like in  \eqref{freeexpt}. Higher order contributions to the Lagrangian should give higher order terms in the expansion \eqref{freeexpt}. One would expect each power of $(l_sT)^4$ to be multiplied by a function of $\lambda$, coming from open string loop corrections (wordsheets with additional holes). It would be interesting to compute these terms and compare the resulting series between weak and strong coupling.

The above discussion naturally raises the question whether it could be that one can generate the full open string theory by an irrelevant deformation of the low energy $\NN=4$ SYM. At first sight this seems unlikely, due to the usual problems with flowing up the renormalization group, but the recent work on $T\bar T$ deformed CFT \cite{Smirnov:2016lqw,Cavaglia:2016oda}, and in particular its single-trace version \cite{Giveon:2017nie,Giveon:2017myj,Asrat:2017tzd,Giribet:2017imm,Chakraborty:2018kpr,Chakraborty:2018aji}, suggest that perhaps this is too pessimistic. In that case, one starts from the infrared $CFT_2$ dual to the near-horizon geometry of a system of strings and fivebranes, perturbs it by an irrelevant, dimension $(2,2)$ operator, and flows in the UV to a theory with a Hagedorn spectrum. There is some evidence that this flow can be made sense of using just the infrared CFT.

In fact, the two systems have some superficial resemblances. In the two-dimensional system, the infrared $CFT_2$ is dual to an $AdS_3$ with $R_{AdS}\sim \sqrt{k}l_s$ (with $k$ the number of fivebranes, taken to be large in the gravity approximation),
and the inverse Hagedorn temperature of the UV theory is $\beta_H\sim R_{AdS}$.
In our case, the infrared $CFT_4$ is dual to an $AdS_5$ with $R_{AdS}\sim \lambda^{\frac{1}{4}}l_s$, and the inverse Hagedorn temperature of the UV theory is again $\beta_H\sim R_{AdS}$.

Thus, it is conceivable that one can understand the free energy of the D3-brane worldvolume theory by starting with the infrared $\NN=4$ SYM and perturbing it as described above. Needless to say, it would be interesting to do that, but it will require new ideas, perhaps along the lines of \cite{Caetano:2020ofu}.

Our discussion raises many additional questions. For example,
\begin{enumerate}
\item It would be interesting to compute other observables in string theory in the background with $N$ D3-branes in the limit \eqref{declim}. A natural class of observables is correlation functions of closed string vertex operators, which should contain information about correlation functions of off-shell operators in the D3-brane worldvolume theory. The analogous problem in single-trace $T\bar T$ deformed CFT was discussed in \cite{Asrat:2017tzd,Giribet:2017imm}.
\item  Our discussion involved the leading order in the $1/N$ expansion. It would be interesting to understand the form of $1/N$ corrections, and in particular the way that the dynamics of closed strings in the bulk comes back when we include these corrections. This might be related to the ideas of Sen on open-closed string duality \cite{Sen:2003xs}.
\item In the limit \eqref{declim}, the non-trivial dynamics involves the open strings. It might be interesting to study the theory in this limit using open string field theory (OSFT). OSFT was found to be useful in some problems involving open string tachyon condensation \cite{Sen:2004nf}, and some other contexts, \eg\ \cite{Sen:2019qqg}. It would be interesting if it could shed light on the physics discussed here.
\item In addition to studying the D3-brane dynamics in the limit \eqref{declim}, we also discussed a deformation of string theory on $AdS_5\times S^5$ which can be thought of from the bulk point of view as studying the geometry \eqref{d3}, \eqref{harfunneg}, and from the boundary point of view as adding to the Lagrangian of $\NN=4$ SYM the dimension eight operator that preserves $\NN=4$ SUSY discussed in the introduction with negative coupling. From the bulk point of view, this theory looks sick. In particular, the metric has a singularity at finite $r$, beyond which it becomes complex. Nevertheless, its thermodynamics studied here looks remarkably well behaved. It would be interesting to study this theory (and its two-dimensional cousin \cite{toappear}) further.
\end{enumerate}

\bigskip

\section*{Acknowledgements}
We thank O. Aharony and A. Hashimoto for discussions, and O. Aharony for comments on the manuscript.
The work of SC is supported by the Infosys Endowment for the study of the Quantum Structure of Spacetime.
The work of AG and DK is supported in part by BSF grant number 2018068.
The work of AG is also supported in part by a center of excellence supported
by the Israel Science Foundation (grant number 2289/18).
The work of DK is also supported in part by DOE grant de-sc0009924.

\bigskip

\appendix
\section{$T(r_0-r_*)$, $C(r_0-r_*)$ and $S(E)$}

In this appendix,
we present the expansions of the temperature $T$ and specific heat $C$ in $r_0-r_*$,
and the behavior of the entropy as a function of energy in various limits.

In section \ref{sec2}, we showed that the temperature is bounded from above and it attains its maximum at $r_0=r_\ast$, \eqref{rstar}.
Near $r_0=r_\ast$, the temperature has the following expansion:
\begin{eqnarray}\label{texp}
T=T_\ast-\frac{3^{7/8}(r_0-r_\ast)^2}{2^{11/4}\pi^{7/4}\lambda^{3/4} l_s^3}+\frac{(r_0-r_\ast)^3}{4\sqrt{2}\pi^{2}\lambda l_s^4}+O\left((r_0-r_\ast)^4\right).
\end{eqnarray}
Figure \ref{img5} shows that near $r_0=r_\ast$, the specific heat diverges.
Around $r_0=r_\ast$, the specific heat behaves as
\begin{eqnarray}\label{spheat}
C=N^2V_3T\left(\partial\SS\over\partial T\right)=-N^2V_3T\left(\partial^2\FF\over\partial T^2\right)=
-\frac{2N^2V_3}{3^{5/4}\pi^{3/2}\sqrt{\lambda}l_s^2 (r_0-r_\ast)}+\dots,
\end{eqnarray}
where $\SS$ and $\FF$ are defined in \eqref{deffff}. Plugging \eqref{texp} into \eqref{spheat}, we see that the specific heat diverges like $C\sim\Delta T^{-\frac{1}{2}}$, in agreement with \eqref{fengexp}.

At large $N$, $S=N^2V_3\SS(\EE\equiv E/N^2V_3)$. The behavior of the entropy density at small energy is
\begin{eqnarray}\label{entdpexp}
\SS&=&\frac{2^{5/4}\sqrt{\pi}}{3^{3/4}}\EE^{3/4}\Bigg{(}1+\frac{4}{3}\left(\pi^{3}\lambda l_s^4 \EE\right)-\frac{8}{27}\left(\pi^{3}\lambda l_s^4 \EE\right)^2+\dots\Bigg{)}.
\end{eqnarray}
The leading term is the entropy density of $3+1$ dimensional $\NN=4$ $U(N)$ SYM, and the higher order terms in $l_s^4\EE$ can be thought of as due to the dimension eight irrelevant deformation discussed in the text.

If one formally continues $r_0$ past $r_*$, one gets for large $\EE$ the micro-canonical entropy
\begin{eqnarray}
 \SS=\frac{2^{15/4}\pi^{2}l_s^2\sqrt{\lambda}}{5^{5/4}}\EE^{5/4}+\dots,
 \end{eqnarray}
which is, of course, the property of a large Schwarzschild black hole in $6+1$ dimensions. As discussed in section 2, such large black holes do not play a role in the thermodynamics. At the same temperature as that of these large black holes, there is a black hole with $r_0<r_*$,
that has smaller free energy, which dominates the thermodynamics. This is consistent with the fact that the large black holes with entropy $\sim E^{5/4}$
have nothing to do with the D3-branes, and therefore cannot play a role in the thermodynamics of their worldvolume theory,
which is governed by the small black holes. In the actual worldvolume theory, the energy density is in fact bounded from above, \eqref{emax}.

In section \ref{sec3}, we discussed the thermodynamics of the {\it negative} D3 black branes.
In this case, the small energy expansion of the entropy density is
\begin{eqnarray}\label{entdpexn}
\SS&=&\frac{2^{5/4}\sqrt{\pi}}{3^{3/4}}\EE^{3/4}\Bigg{(}1-\frac{4}{3}\left(\pi^{3}\lambda l_s^4 \EE\right)-\frac{8}{27}\left(\pi^{3}\lambda l_s^4 \EE\right)^2+\dots\Bigg{)}.
\end{eqnarray}
Again, the leading term is the entropy of ${\cal N}=4$ SYM,
and the higher order terms in $l_s^4\EE$ are due to its irrelevant deformation.
But, now, the dimension eight deformation is with a {\it negative} coupling, $-l_s^4$ (compare \eqref{entdpexn} to \eqref{entdpexp}).
Finally, the large energy behavior of the entropy is given in \eqref{entbhflatn}.

\newpage

%\bibliography{ref}\bibliographystyle{JHEP}

\begin{thebibliography}{10}

\bibitem{Maldacena:1997re}
J.~M. Maldacena, \emph{{The Large N limit of superconformal field theories and
  supergravity}}, \href{https://doi.org/10.1023/A:1026654312961}{\emph{Int. J.
  Theor. Phys.} {\bfseries 38} (1999) 1113}
  [\href{https://arxiv.org/abs/hep-th/9711200}{{\ttfamily hep-th/9711200}}].

\bibitem{Gubser:1998bc}
S.~Gubser, I.~R. Klebanov and A.~M. Polyakov, \emph{{Gauge theory correlators
  from noncritical string theory}},
  \href{https://doi.org/10.1016/S0370-2693(98)00377-3}{\emph{Phys. Lett. B}
  {\bfseries 428} (1998) 105}
  [\href{https://arxiv.org/abs/hep-th/9802109}{{\ttfamily hep-th/9802109}}].

\bibitem{Witten:1998qj}
E.~Witten, \emph{{Anti-de Sitter space and holography}},
  \href{https://doi.org/10.4310/ATMP.1998.v2.n2.a2}{\emph{Adv. Theor. Math.
  Phys.} {\bfseries 2} (1998) 253}
  [\href{https://arxiv.org/abs/hep-th/9802150}{{\ttfamily hep-th/9802150}}].

\bibitem{Aharony:1999ti}
O.~Aharony, S.~S. Gubser, J.~M. Maldacena, H.~Ooguri and Y.~Oz, \emph{{Large N
  field theories, string theory and gravity}},
  \href{https://doi.org/10.1016/S0370-1573(99)00083-6}{\emph{Phys. Rept.}
  {\bfseries 323} (2000) 183}
  [\href{https://arxiv.org/abs/hep-th/9905111}{{\ttfamily hep-th/9905111}}].

\bibitem{Gubser:1998iu}
S.~S. Gubser and A.~Hashimoto, \emph{{Exact absorption probabilities for the
  D3-brane}}, \href{https://doi.org/10.1007/s002200050614}{\emph{Commun. Math.
  Phys.} {\bfseries 203} (1999) 325}
  [\href{https://arxiv.org/abs/hep-th/9805140}{{\ttfamily hep-th/9805140}}].

\bibitem{Danielsson:2000ze}
U.~H. Danielsson, A.~Guijosa, M.~Kruczenski and B.~Sundborg, \emph{{D3-brane
  holography}},
  \href{https://doi.org/10.1088/1126-6708/2000/05/028}{\emph{JHEP} {\bfseries
  05} (2000) 028} [\href{https://arxiv.org/abs/hep-th/0004187}{{\ttfamily
  hep-th/0004187}}].

\bibitem{Kutasov:2001uf}
D.~Kutasov, \emph{{Introduction to little string theory}}, {\emph{ICTP Lect.
  Notes Ser.} {\bfseries 7} (2002) 165}.

\bibitem{Intriligator:1999ai}
K.~A. Intriligator, \emph{{Maximally supersymmetric RG flows and AdS duality}},
  \href{https://doi.org/10.1016/S0550-3213(99)00803-2}{\emph{Nucl. Phys. B}
  {\bfseries 580} (2000) 99}
  [\href{https://arxiv.org/abs/hep-th/9909082}{{\ttfamily hep-th/9909082}}].

\bibitem{Gubser:1997yh}
S.~S. Gubser, I.~R. Klebanov and A.~A. Tseytlin, \emph{{String theory and
  classical absorption by three-branes}},
  \href{https://doi.org/10.1016/S0550-3213(97)00325-8}{\emph{Nucl. Phys. B}
  {\bfseries 499} (1997) 217}
  [\href{https://arxiv.org/abs/hep-th/9703040}{{\ttfamily hep-th/9703040}}].

\bibitem{Gubser:1997se}
S.~S. Gubser and I.~R. Klebanov, \emph{{Absorption by branes and Schwinger
  terms in the world volume theory}},
  \href{https://doi.org/10.1016/S0370-2693(97)01099-X}{\emph{Phys. Lett. B}
  {\bfseries 413} (1997) 41}
  [\href{https://arxiv.org/abs/hep-th/9708005}{{\ttfamily hep-th/9708005}}].

\bibitem{Gubser:1998kv}
S.~S. Gubser, A.~Hashimoto, I.~R. Klebanov and M.~Krasnitz, \emph{{Scalar
  absorption and the breaking of the world volume conformal invariance}},
  \href{https://doi.org/10.1016/S0550-3213(98)00301-0}{\emph{Nucl. Phys. B}
  {\bfseries 526} (1998) 393}
  [\href{https://arxiv.org/abs/hep-th/9803023}{{\ttfamily hep-th/9803023}}].

\bibitem{Costa:1999sk}
M.~S. Costa, \emph{{Absorption by double centered D3-branes and the Coulomb
  branch of N=4 SYM theory}},
  \href{https://doi.org/10.1088/1126-6708/2000/05/041}{\emph{JHEP} {\bfseries
  05} (2000) 041} [\href{https://arxiv.org/abs/hep-th/9912073}{{\ttfamily
  hep-th/9912073}}].

\bibitem{Costa:2000gk}
M.~S. Costa, \emph{{A Test of the AdS / CFT duality on the Coulomb branch}},
  \href{https://doi.org/10.1016/S0370-2693(00)00484-6}{\emph{Phys. Lett. B}
  {\bfseries 482} (2000) 287}
  [\href{https://arxiv.org/abs/hep-th/0003289}{{\ttfamily hep-th/0003289}}].

\bibitem{Rastelli:2000xj}
L.~Rastelli and M.~Van~Raamsdonk, \emph{{A Note on dilaton absorption and near
  infrared D3-brane holography}},
  \href{https://doi.org/10.1088/1126-6708/2000/12/005}{\emph{JHEP} {\bfseries
  12} (2000) 005} [\href{https://arxiv.org/abs/hep-th/0011044}{{\ttfamily
  hep-th/0011044}}].

\bibitem{Evans:2001zn}
N.~J. Evans, C.~V. Johnson and M.~Petrini, \emph{{Clearing the throat:
  Irrelevant operators and finite temperature in large N gauge theory}},
  \href{https://doi.org/10.1088/1126-6708/2002/05/002}{\emph{JHEP} {\bfseries
  05} (2002) 002} [\href{https://arxiv.org/abs/hep-th/0112058}{{\ttfamily
  hep-th/0112058}}].

\bibitem{Niarchos:2017cdz}
V.~Niarchos, \emph{{Holographic entanglement entropy in open-closed string
  duality}},  \href{https://arxiv.org/abs/1701.03113}{{\ttfamily 1701.03113}}.

\bibitem{Ferko:2019oyv}
C.~Ferko, H.~Jiang, S.~Sethi and G.~Tartaglino-Mazzucchelli, \emph{{Non-linear
  supersymmetry and $ T\overline{T} $-like flows}},
  \href{https://doi.org/10.1007/JHEP02(2020)016}{\emph{JHEP} {\bfseries 02}
  (2020) 016} [\href{https://arxiv.org/abs/1910.01599}{{\ttfamily
  1910.01599}}].

\bibitem{Caetano:2020ofu}
J.~Caetano, W.~Peelaers and L.~Rastelli, \emph{{Maximally Supersymmetric RG
  Flows in 4D and Integrability}},
  \href{https://arxiv.org/abs/2006.04792}{{\ttfamily 2006.04792}}.

\bibitem{Giveon:2017nie}
A.~Giveon, N.~Itzhaki and D.~Kutasov, \emph{{$ \mathrm{T}\overline{\mathrm{T}}
  $ and LST}}, \href{https://doi.org/10.1007/JHEP07(2017)122}{\emph{JHEP}
  {\bfseries 07} (2017) 122}
  [\href{https://arxiv.org/abs/1701.05576}{{\ttfamily 1701.05576}}].

\bibitem{Giveon:2017myj}
A.~Giveon, N.~Itzhaki and D.~Kutasov, \emph{{A solvable irrelevant deformation
  of AdS$_{3}$/CFT$_{2}$}},
  \href{https://doi.org/10.1007/JHEP12(2017)155}{\emph{JHEP} {\bfseries 12}
  (2017) 155} [\href{https://arxiv.org/abs/1707.05800}{{\ttfamily
  1707.05800}}].

\bibitem{Asrat:2017tzd}
M.~Asrat, A.~Giveon, N.~Itzhaki and D.~Kutasov, \emph{{Holography Beyond AdS}},
  \href{https://doi.org/10.1016/j.nuclphysb.2018.05.005}{\emph{Nucl. Phys. B}
  {\bfseries 932} (2018) 241}
  [\href{https://arxiv.org/abs/1711.02690}{{\ttfamily 1711.02690}}].

\bibitem{Giribet:2017imm}
G.~Giribet, \emph{{$T\bar{T}$-deformations, AdS/CFT and correlation
  functions}}, \href{https://doi.org/10.1007/JHEP02(2018)114}{\emph{JHEP}
  {\bfseries 02} (2018) 114}
  [\href{https://arxiv.org/abs/1711.02716}{{\ttfamily 1711.02716}}].

\bibitem{Chakraborty:2018kpr}
S.~Chakraborty, A.~Giveon, N.~Itzhaki and D.~Kutasov, \emph{{Entanglement
  beyond AdS}},
  \href{https://doi.org/10.1016/j.nuclphysb.2018.08.011}{\emph{Nucl. Phys. B}
  {\bfseries 935} (2018) 290}
  [\href{https://arxiv.org/abs/1805.06286}{{\ttfamily 1805.06286}}].

\bibitem{Chakraborty:2018aji}
S.~Chakraborty, \emph{{Wilson loop in a $T\bar{T}$ like deformed
  $\rm{CFT}_2$}},
  \href{https://doi.org/10.1016/j.nuclphysb.2018.12.003}{\emph{Nucl. Phys. B}
  {\bfseries 938} (2019) 605}
  [\href{https://arxiv.org/abs/1809.01915}{{\ttfamily 1809.01915}}].

\bibitem{Kutasov:1999xu}
D.~Kutasov and N.~Seiberg, \emph{{More comments on string theory on AdS(3)}},
  \href{https://doi.org/10.1088/1126-6708/1999/04/008}{\emph{JHEP} {\bfseries
  04} (1999) 008} [\href{https://arxiv.org/abs/hep-th/9903219}{{\ttfamily
  hep-th/9903219}}].

\bibitem{Aharony:1998ub}
O.~Aharony, M.~Berkooz, D.~Kutasov and N.~Seiberg, \emph{{Linear dilatons, NS
  five-branes and holography}},
  \href{https://doi.org/10.1088/1126-6708/1998/10/004}{\emph{JHEP} {\bfseries
  10} (1998) 004} [\href{https://arxiv.org/abs/hep-th/9808149}{{\ttfamily
  hep-th/9808149}}].

\bibitem{Aharony:1999ks}
O.~Aharony, \emph{{A Brief review of 'little string theories'}},
  \href{https://doi.org/10.1088/0264-9381/17/5/302}{\emph{Class. Quant. Grav.}
  {\bfseries 17} (2000) 929}
  [\href{https://arxiv.org/abs/hep-th/9911147}{{\ttfamily hep-th/9911147}}].

\bibitem{Aharony:2004xn}
O.~Aharony, A.~Giveon and D.~Kutasov, \emph{{LSZ in LST}},
  \href{https://doi.org/10.1016/j.nuclphysb.2004.05.015}{\emph{Nucl. Phys. B}
  {\bfseries 691} (2004) 3}
  [\href{https://arxiv.org/abs/hep-th/0404016}{{\ttfamily hep-th/0404016}}].

\bibitem{toappear}
S.~Chakraborty, A.~Giveon and D.~Kutasov, \emph{{$T\bar{T}$, Black Holes and
  Negative Strings}},  \href{https://arxiv.org/abs/2006.13249}{{\ttfamily
  2006.13249}}.

\bibitem{Dijkgraaf:2016lym}
R.~Dijkgraaf, B.~Heidenreich, P.~Jefferson and C.~Vafa, \emph{{Negative Branes,
  Supergroups and the Signature of Spacetime}},
  \href{https://doi.org/10.1007/JHEP02(2018)050}{\emph{JHEP} {\bfseries 02}
  (2018) 050} [\href{https://arxiv.org/abs/1603.05665}{{\ttfamily
  1603.05665}}].

\bibitem{Horowitz:1991cd}
G.~T. Horowitz and A.~Strominger, \emph{{Black strings and P-branes}},
  \href{https://doi.org/10.1016/0550-3213(91)90440-9}{\emph{Nucl. Phys. B}
  {\bfseries 360} (1991) 197}.

\bibitem{Smirnov:2016lqw}
F.~Smirnov and A.~Zamolodchikov, \emph{{On space of integrable quantum field
  theories}},
  \href{https://doi.org/10.1016/j.nuclphysb.2016.12.014}{\emph{Nucl. Phys. B}
  {\bfseries 915} (2017) 363}
  [\href{https://arxiv.org/abs/1608.05499}{{\ttfamily 1608.05499}}].

\bibitem{Cavaglia:2016oda}
A.~Cavaglià, S.~Negro, I.~M. Szécsényi and R.~Tateo, \emph{{$T
  \bar{T}$-deformed 2D Quantum Field Theories}},
  \href{https://doi.org/10.1007/JHEP10(2016)112}{\emph{JHEP} {\bfseries 10}
  (2016) 112} [\href{https://arxiv.org/abs/1608.05534}{{\ttfamily
  1608.05534}}].

\bibitem{Aharony:2018bad}
O.~Aharony, S.~Datta, A.~Giveon, Y.~Jiang and D.~Kutasov, \emph{{Modular
  invariance and uniqueness of $T\bar{T}$ deformed CFT}},
  \href{https://doi.org/10.1007/JHEP01(2019)086}{\emph{JHEP} {\bfseries 01}
  (2019) 086} [\href{https://arxiv.org/abs/1808.02492}{{\ttfamily
  1808.02492}}].

\bibitem{Gubser:1996de}
S.~Gubser, I.~R. Klebanov and A.~Peet, \emph{{Entropy and temperature of black
  3-branes}}, \href{https://doi.org/10.1103/PhysRevD.54.3915}{\emph{Phys. Rev.
  D} {\bfseries 54} (1996) 3915}
  [\href{https://arxiv.org/abs/hep-th/9602135}{{\ttfamily hep-th/9602135}}].

\bibitem{Sen:2003xs}
A.~Sen, \emph{{Open closed duality at tree level}},
  \href{https://doi.org/10.1103/PhysRevLett.91.181601}{\emph{Phys. Rev. Lett.}
  {\bfseries 91} (2003) 181601}
  [\href{https://arxiv.org/abs/hep-th/0306137}{{\ttfamily hep-th/0306137}}].

\bibitem{Sen:2004nf}
A.~Sen, \emph{{Tachyon dynamics in open string theory}},
  \href{https://doi.org/10.1142/S0217751X0502519X}{\emph{Int. J. Mod. Phys. A}
  {\bfseries 20} (2005) 5513}
  [\href{https://arxiv.org/abs/hep-th/0410103}{{\ttfamily hep-th/0410103}}].

\bibitem{Sen:2019qqg}
A.~Sen, \emph{{Fixing an Ambiguity in Two Dimensional String Theory Using
  String Field Theory}},
  \href{https://doi.org/10.1007/JHEP03(2020)005}{\emph{JHEP} {\bfseries 03}
  (2020) 005} [\href{https://arxiv.org/abs/1908.02782}{{\ttfamily
  1908.02782}}].

\end{thebibliography}

\providecommand{\href}[2]{#2}\begingroup\raggedright\endgroup

\end{document}